\documentclass[aip,reprint,apm]{revtex4-2}

\usepackage[separate-uncertainty=true,multi-part-units=single]{siunitx}
\usepackage{chemformula}
\usepackage{amssymb}
\newcommand{\mytilde}{\raise.17ex\hbox{$\scriptstyle\mathtt{\sim}$}}
\newcommand{\layer}[2]{\allowbreak\ch{#1}(#2)}
\newcommand{\supsec}[1]{\textrm{Supplementary Section #1}}
\newcommand{\supfig}[1]{\textrm{Supplementary Figure #1}}

\draft

\begin{document}

\title{The multi-scale dynamics of all-optical exchange bias reversal}

\author{Floris J. F. van Riel}
\email[]{f.j.f.v.riel@tue.nl}
\author{Andries B. M. Droste}
\author{Bert Koopmans}
\author{Diana C. Leitao}
\affiliation{Department of Applied Physics, Eindhoven University of Technology, P.O. Box 513, 5600 MB Eindhoven, The Netherlands}

\date{\today}

\begin{abstract}
Pinning magnetization in a ferromagnetic thin film is commonly realized through exchange biasing with an adjacent antiferromagnet. Field-cooling from above the N\'{e}el temperature is a reliable yet slow re-pinning method in exchange-biased systems. For on-demand reprogrammable devices, localized and rapid exchange bias repinning methods are essential. Recent work has shown that femtosecond laser pulses enable field-free reversal of exchange bias in tailored multilayer stacks. Contrary to field-cooling, our experiments with ultrafast excitation reach hitherto unexplored regimes in the exchange bias setting process. Here, we unravel these observations by considering both ultrafast magnetization dynamics on the femto- to picosecond timescale and slow heat-driven dynamics on millisecond timescales and upwards. We develop a microscopic framework of exchange bias setting in a polycrystalline antiferromagnetic thin film like \ch{IrMn} that provides a complete description of the observations in our present experiments and those found in literature. We expand the use of our model by identifying material platforms and stack designs that lead to optimized performance, aiding further development of optically reprogrammable devices.
\end{abstract}

\maketitle

\section{Introduction}
In spintronics devices, exchange bias is frequently used as a robust magnetic pinning mechanism that induces a preferential magnetization orientation at a ferromagnetic/antiferromagnetic interface \cite{Dieny2020}. This phenomenon enables stable referencing solutions, e.g., for magnetic sensors \cite{Leitao2024} or memory devices \cite{Guo2022a}. Iridium-manganese alloys \ch{Ir_{x}Mn_{1-x}} are a type of antiferromagnet that is commonly used in sputter-deposited thin film stacks for exchange biasing transition-metal ferromagnetic systems\cite{Sort2005,Chen2014,Feng2017,Gao2019,Guo2022,Wang2023}. When depositing \ch{IrMn} on a ferromagnetic layer, its magnetic configuration will be imprinted in the \ch{IrMn}, imposing an effective unidirectional exchange bias field $H_{\mathrm{EB}}$ on the ferromagnet via exchange coupling with the \ch{Mn} spins at the interface \cite{Meiklejohn1956,Nogues1999}. After deposition, manipulation of the exchange bias may be realized by field-cooling through the N\'{e}el temperature, either globally by thermal annealing or locally through Joule heating \cite{Cao2010,Papusoi2008,Wang2024} or laser heating \cite{Berthold2014,Almeida2015,Ueberschar2015b}.

These annealing methods rely on an externally applied magnetic field to guide the setting of $H_{\mathrm{EB}}$. However, recent research has shown the possibility of field-free and ultrafast manipulation of exchange bias by all-optical magnetization switching with femtosecond laser pulses \cite{Vallobra2017,Guo2024,VanRiel2025}. Exchange bias reversal with single-shot laser pulses was first observed in thin film stacks containing antiferromagnetic \ch{IrMn} and \ch{Co}-\ch{Gd} ferrimagnetic alloys \cite{Guo2024} or multilayers \cite{VanRiel2025}. In our recent study \cite{VanRiel2025}, we showed how specifically tailored thin film stacks experience a full \SI{180}{\degree} reversal of $H_{\mathrm{EB}}$ from up to down and vice versa upon excitation by a single laser pulse. Even though the reversal effect itself is evident, our present results also indicate a creep-like evolution of the magnitude of $H_{\mathrm{EB}}$ across several days that hitherto remained unexplored. Particularly, grasping the complete picture of the underlying mechanism is challenged by the complexity of the setting process occurring because of the ultrafast heating and the highly non-equilibrium states governing the dynamics, but these processes are critical for determining the long-term pinning stability in devices.

In this work, we show how a mixture of ultrafast excitation and slow temperature relaxation can explain the creep of $H_{\mathrm{EB}}$ and how future applied research may leverage these concepts for optimized performance in reprogrammable devices. We present a model based on previous works \cite{Vallejo-Fernandez2008,Kanso2019,Khamtawi2023,Jenkins2021} that in all its simplicity produces very rich and unexpected results when combined with ultrafast magnetization switching. A wide range of timescales (from \SIrange[range-units=single,range-phrase=\text{ to }]{e-15}{e17}{\second}) is considered to explain observations from present experiments and reveal the multi-scale dynamics of all-optical exchange bias reversal. The outcomes provide guidelines for future research on designing and fabricating laser-reprogrammable exchange-biased devices.
\section{Experimental demonstration}
\begin{figure}%
	\centering%
	\includegraphics[width=\columnwidth]{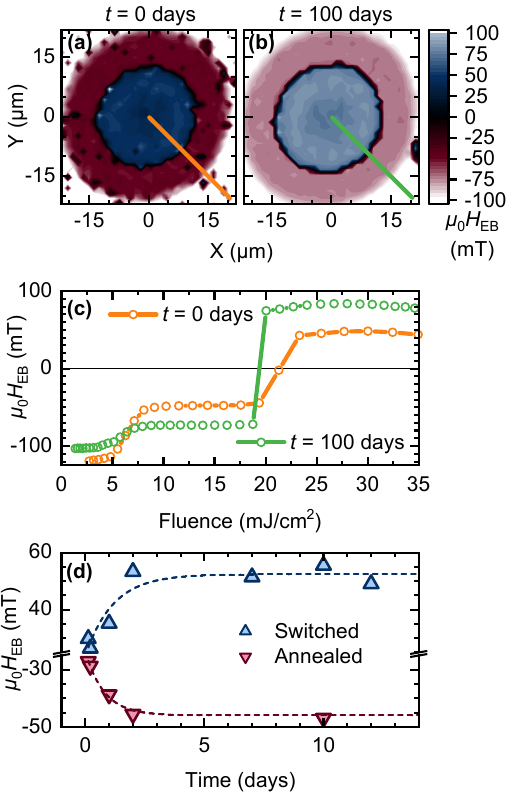}%
	\caption{\label{fig:1}(a) Experimentally measured map of the exchange bias field magnitude $H_{\mathrm{EB}}$ across a region of the sample that was illuminated by a single femtosecond Gaussian laser spot with a peak fluence of \SI{38}{\milli\joule\per\square\centi\meter}. The measurement is performed by Kerr microscopy field sweeps directly after laser excitation, where $H_{\mathrm{EB}}$ is locally extracted from the hysteresis curve field shift. (b) The same measurement as in (a) but performed \SI{100}{days} after illumination. (c) Radial cross-sections of the distribution of $H_{\mathrm{EB}}$ as indicated by the orange and green lines in (a) and (b). (d) A demonstration of the evolution of $H_{\mathrm{EB}}$ from a newly deposited and switched sample over the period of two weeks, for both the annealed (red) and switched (blue) regions from (a). The dashed lines are guides to the eye.}%
\end{figure}%
The creep-like evolution of $H_{\mathrm{EB}}$ was observed in thin film stacks from our recent study \cite{VanRiel2025} \layer{Ta}{4}/\layer{Pt}{4}/\layer{Co}{0.6}/\layer{Gd}{5.5}/\layer{Co}{1}/[\layer{Pt}{1.25}/\layer{Co}{0.6}]$_{\mathrm{x}2}$/\layer{Ir_{20}Mn_{80}}{4}/\layer{Ta}{4} (thicknesses in \si{\nano\meter}) grown by magnetron sputtering on a thermally oxidized \ch{Si}/\ch{SiO2} wafer. On these stacks, we measure the magnetic response to excitation by a \mytilde\SI{100}{\femto\second} linearly polarized laser pulse with a Gaussian spatial energy profile and a central wavelength of \SI{700}{\nano\meter}. Hysteresis loops were measured by \textit{ex-situ} Kerr microscopy of the illuminated sample area. The measurement was performed once directly after excitation at $t_{\mathrm{meas}}=\SI{0}{days}$ and then a second time after approximately $t_{\mathrm{meas}}=\SI{100}{days}$, at which point a significantly different magnitude of $H_{\mathrm{EB}}$ is measured, which is shown in Fig.~\ref{fig:1}. Figure~\ref{fig:1}a shows the top-down processed microscopy image from the initial measurement and Fig.~\ref{fig:1}b shows the measurement after \SI{100}{days}, with blue and red colors indicating positive and negative values of $H_{\mathrm{EB}}$. The Gaussian fluence (energy per area) profile of the laser pulse is reflected in the circularly symmetric profile of $H_{\mathrm{EB}}$. The color map reveals three general regions: unaffected for low fluences (white), affected but not switched (annealed) for medium fluences (red) and switched for high fluences (blue). By comparing the colors between Fig.~\ref{fig:1}a and Fig.~\ref{fig:1}b it is evident that the profile of the annealed/switched regions is identical, but the absolute values of $H_{\mathrm{EB}}$ have increased in both regions. The orange and green lines are cross-sections of the profile that are plotted in Fig.~\ref{fig:1}c. Similar trends of $H_{\mathrm{EB}}$ evolving over several days at room temperature have been reported previously by Migliorini et al. \cite{Migliorini2018} for polycrystalline \ch{IrMn} films, where it is also speculated that direct exchange between spins at the ferromagnetic/antiferromagnetic interface drives the setting process.

We also studied the time-dependence of $H_{\mathrm{EB}}$ on the timescale of days and found that both the annealed and switched regions evolve with approximately the same timescale, as shown in Fig.~\ref{fig:1}d. To capture the complete picture we formulated a model\cite{VanRiel2025} that extends the magnetization dynamics beyond the ultrafast regime. We identified thermally assisted relaxation of antiferromagnetic grains as the primary mechanism for explaining the slow creep-like evolution of $H_{\mathrm{EB}}$. Below we elaborate on the extended models for both the ultrafast and slow magnetization dynamics and how they are linked together to describe the full exchange bias reversal process from excitation to relaxation.

\section{Modeling the reversal process}
\begin{figure*}%
	\centering%
	\includegraphics[width=\textwidth]{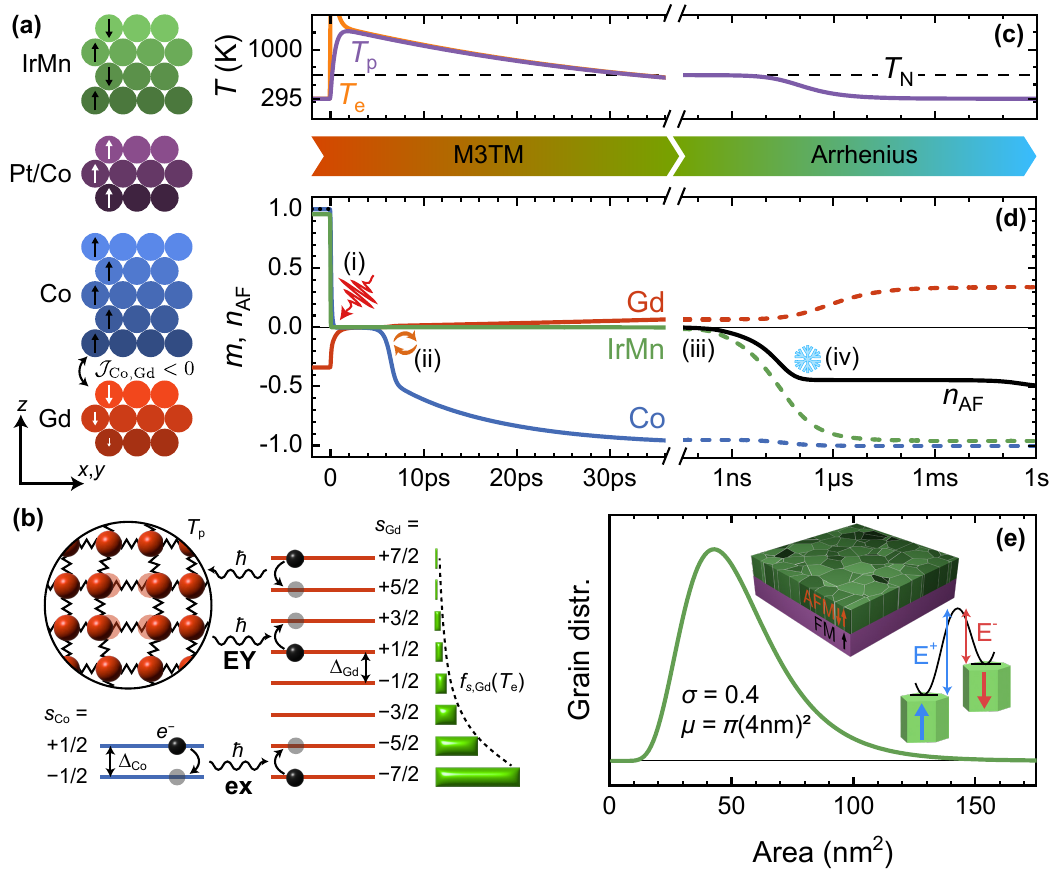}%
	\caption{\label{fig:2}(a) Overview of the simulated stack, consisting of in total 15 atomic monolayers of \ch{Gd}, \ch{Co}, \ch{Pt/Co} (modeled as a single compound ferromagnet) and \ch{IrMn} (modeled as a layered antiferromagnet). The symbol $\mathcal{J}_{ij}$ represents the exchange coupling between atoms of species $i$ and $j$. (b) Schematic of the processes modeled by the M3TM framework, exemplified for \ch{Gd} and \ch{Co}. The lines represent the energy levels for the various spin quantum numbers $s_{i}$ which are spaced by the exchange splitting $\Delta_{i}$. On top is shown the Elliot-Yafet (EY) spin-flip scattering by creation/annihilation of a phonon and on the bottom is shown the exchange scattering (ex) between two electrons exchanging angular momentum. The green bars on the right represent the occupation fraction $f_{s,\ch{Gd}}$ of each spin level in \ch{Gd} in equilibrium. (c) The evolution of temperature $T$ over time for illumination by a \SI{58.3e8}{\joule\per\cubic\meter} laser pulse, also taking into account the difference between the electron ($T_{\mathrm{e}}$) and phonon ($T_{\mathrm{p}}$) temperatures within the first picosecond after excitation. The dashed line represents the N\'{e}el temperature $T_{\mathrm{N}}$. (d) Evolution of the layer-averaged magnetization $m$ in the \ch{Gd} (red), \ch{Co} (blue) and \ch{IrMn} (green) subsystems, alongside de exchange bias parameter $n_{\mathrm{AF}}$. The dotted line before $0$ and the dashed lines after \mytilde\SI{36}{\pico\second} represent their respective values in thermal equilibrium. Above the plot is indicated which modeling framework \textendash{} M3TM or Arrhenius \textendash{} is used in each temporal regime. Four key events are highlighted in roman numerals: (i) laser incidence, (ii) ferromagnetic reversal and remagnetization, (iii) $T$ dropping below $T_{\mathrm{N}}$ and antiferromagnetic remagnetization and (iv) freezing in of the exchange bias. (e) Plot of the log-normally distributed areas of the grains, with parameters $\mu$ and $\sigma$ as used in Eq.~\ref{eq:lognormal}. The inset illustrates how the Arrhenius law is applied to the antiferromagnet (AFM). When the ferromagnet (FM) is uniformly magnetized up, the grains that contribute positively to the exchange bias have a lower energy than grains that contribute negatively. The energy required for a grain to spontaneously transition from up to down (down to up) is indicated by $E^{+}$ ($E^{-}$).}%
\end{figure*}%
In order for our model to predict the behavior of the exchange bias upon ultrafast excitation, it first needs to predict the outcome of the magnetization switching in the ferrimagnet for a given laser fluence $P_{0}$. This is achieved using the layered microscopic three temperature model (M3TM) \cite{Koopmans2010,Beens2019}, which models magnetization dynamics on the femtosecond to picosecond timescale. We model our stacks by the layered system shown in Fig.~\ref{fig:2}a. Here, a `layer' is an atomically thin entity with an associated normalized magnetization $m_{z}=M_{z}/M(T=\SI{0}{\kelvin})$ along the perpendicular $z$-axis, arising from the spins of the electrons. We consider a stack with perpendicular magnetic anisotropy so only the $z$-component of $m$ is non-zero, i.e. $m=m_{z}$. Magnetic exchange interaction between atoms \textendash{} both within and between layers \textendash{} is captured by an exchange splitting energy $\Delta_{i}$ separating spin energy levels, given by Eq.~\ref{eq:delta} and schematically illustrated in Fig.~\ref{fig:2}b. We use spin quantum numbers $S_{\mathrm{Co}}=S_{\mathrm{Mn}}=1/2$ and $S_{\mathrm{Gd}}=7/2$ for the respective \ch{Co}, \ch{Mn} and \ch{Gd} atoms \cite{Coey2001}.

The laser heat gets distributed over two thermodynamic systems: electrons directly absorb the laser light and transfer it to phonons via electron-phonon scattering. Both electron and phonon systems are given an associated temperature $T_{\mathrm{e}}$ and $T_{\mathrm{p}}$, respectively, which are coupled via Eq.~\ref{eq:2tm}. Figure~\ref{fig:2}c shows the evolution of $T_{\mathrm{e}}$ and $T_{\mathrm{p}}$ as a function of time after laser excitation, where the excitation itself occurs at time $t=0$ and is annotated by the roman numeral (i) in Fig.~\ref{fig:2}d. The magnetization reversal is driven by Elliot-Yafet electron-phonon spin-flip scattering and by electron-electron angular momentum exchange scattering, as schematically shown in Fig.~\ref{fig:2}b. Both Elliot-Yafet and exchange scattering processes depend on $T_{\mathrm{e}}$ and/or $T_{\mathrm{p}}$ following Eq.~\ref{eq:fsi}. Figure~\ref{fig:2}d shows how the magnetization $m$ evolves over time for a fluence $P_{0}=\SI{58.3e5}{\milli\joule\per\cubic\centi\meter}$ above the switching threshold, noting that the synthetic ferrimagnet \ch{Co}/\ch{Gd} remagnetizes in the opposite direction when the temperature drops below the Curie temperature, as annotated with (ii) in Fig.~\ref{fig:2}d. Details on implementation of the M3TM are given in Appendix~\ref{app:m3tm}.

The outcome of the M3TM is binary, i.e. it can only predict whether or not reversal of the exchange bias takes place by looking at the magnetic state (up or down) of the \ch{Pt}/\ch{Co} system after remagnetization. At this time after excitation, ultrafast processes no longer play a significant role. In this regime, we approximate the system temperature as $T=T_{\mathrm{e}}=T_{\mathrm{p}}$ and we assume the magnetization in all layers is equal to its equilibrium value at temperature $T$ according to Eq.~\ref{eq:equilibrium}. In order to now predict the exchange bias magnitude, we simulate the magnetic state of the antiferromagnet in more detail.

The polycrystalline antiferromagnet is modeled as a collection of non-interacting columnar grains with a fixed distribution (median area $\mu$ and standard deviation $\sigma$, see Fig.~\ref{fig:2}e). Note that when we refer to the grain area, we mean the area $A$ of a particular grain from the distribution and not the median area $\mu$ of the whole distribution. All grains are subject to heat-driven transitions between states separated by an energy barrier. Because of the ultrafast excitation and subsequent rapid cooling, it is expected that grains are excited far out of their thermal equilibrium state and may become trapped in metastable states once the energy barriers exceed the thermal energy required to overcome them. These processes become leading as soon as the temperature drops below the magnetic ordering temperature of the antiferromagnet, that is the N\'{e}el temperature $T_{\mathrm{N}}$, annotated in Fig.~\ref{fig:2}d with (iii).

Transitions under the influence of thermal energy are modeled using Arrhenius' law. Transitions in the ferromagnet may be neglected due to the large intralayer exchange coupling that is much stronger than the interlayer exchange bias coupling. For simplicity, each grain either contributes positively ($\uparrow$) or negatively ($\downarrow$) to the total $H_{\mathrm{EB}}$. We label the fraction of grains with area $A$ that contribute positively (negatively) to $H_{\mathrm{EB}}$ as $n^{\uparrow}$ ($n^{\downarrow}$), and the timescale for spontaneously switching from $\uparrow$ to $\downarrow$ ($\downarrow$ to $\uparrow$) as $\tau_{\uparrow\downarrow}$ ($\tau_{\downarrow\uparrow}$).

Arrhenius' law dictates that these timescales are proportional to the exponential of an energy barrier (see Fig.~\ref{fig:2}e). We will assume that the energy landscape of each antiferromagnetic grain is determined by a uniaxial perpendicular anisotropy and by a unidirectional exchange coupling across the interface with the adjacent ferromagnet. The anisotropy energy density is given by $K_{\mathrm{AF}}=K_{\mathrm{AF},0}m_{i}^{2}$ and the exchange energy per unit area by $\mathcal{J}_{\mathrm{ex}}=\mathcal{J}_{\mathrm{ex},0}m_{i}m_{j}$, where $m_{i}$ and $m_{j}$ are the antiferromagnetic and ferromagnetic equilibrium magnetization according to Eq.~\ref{eq:equilibrium}, respectively. We use $K_{\mathrm{AF},0}=\SI{5.56e5}{\joule\per\cubic\meter}$ for \ch{IrMn} \cite{Khamtawi2023} and an area-dependent $\mathcal{J}_{\mathrm{ex},0}$ as explained in \supsec{1}\cite{Malozemoff1987,Uyama1997,Fuke1999}. We note that the sign of $\mathcal{J}_{\mathrm{ex}}$ is determined by the signs of $m_{i}$ and $m_{j}$. Combining all this into expressions for the transition times we get \cite{Fulcomer1972,Nishioka1996}%
\begin{subequations}\label{eq:transtimes}%
\begin{eqnarray}%
	\tau_{\uparrow\downarrow}&=&\tau_{0}\exp\left(\left.E^{+}\middle/k_{\mathrm{B}}T\right.\right),\\%
	\tau_{\downarrow\uparrow}&=&\tau_{0}\exp\left(\left.E^{-}\middle/k_{\mathrm{B}}T\right.\right),\\%
	E^{\pm}&=&A\begin{cases}%
		K_{\mathrm{AF}}t_{\mathrm{AF}}\pm\mathcal{J}_{\mathrm{ex}}+\left.\mathcal{J}_{\mathrm{ex}}^{2}\middle/4K_{\mathrm{AF}}t_{\mathrm{AF}}\right.\\%
		\qquad\mathrm{if}\left|\mathcal{J}_{\mathrm{ex}}\right|<2K_{\mathrm{AF}}t_{\mathrm{AF}}\\%
		\max\left(0,\pm2\mathcal{J}_{\mathrm{ex}}\right)\\%
		\qquad\mathrm{if}\left|\mathcal{J}_{\mathrm{ex}}\right|>2K_{\mathrm{AF}}t_{\mathrm{AF}}%
	\end{cases},\nonumber%
\end{eqnarray}%
\end{subequations}%
where $\tau_{0}\mathrel{\mytilde}\SI{e-10}{\second}$ is the attempt time. Note that the transition times in Eqs.~\ref{eq:transtimes} are strongly dependent on $T$ (also through $m_{i,j}$), the grain area $A$ and the antiferromagnetic thickness $t_{\mathrm{AF}}$.

To quantify the grain alignment, we define the parameter $n_{\mathrm{AF}}\equiv n^{\uparrow}-n^{\downarrow}$ which takes values in the range $\left[-1,1\right]$, where a value of $+1$ ($-1$) means all grains with area $A$ contribute positively (negatively) to $H_{\mathrm{EB}}$. From Arrhenius' law, an ordinary differential rate equation governing the evolution of $n_{\mathrm{AF}}$ can be derived, and is given by%
\begin{equation}\label{eq:maf}%
	\frac{\mathrm{d}n_{\mathrm{AF}}}{\mathrm{d}t}=\left(\frac{1}{\tau_{\downarrow\uparrow}}-\frac{1}{\tau_{\uparrow\downarrow}}\right)\left[1-n_{\mathrm{AF}}\coth\left(\frac{\mathcal{J}_{\mathrm{ex}}A}{k_{\mathrm{B}}T}\right)\right].%
\end{equation}%
This equation may be solved for all areas $A$ and then integrated over the entire distribution to find the magnitude $h_{\mathrm{EB}}$ of exchange bias according to Eq.~\ref{eq:heb}. This value is normalized to the maximal attainable exchange bias $H_{\mathrm{EB,max}}$ for an infinitely thick antiferromagnet at $T=\SI{0}{\kelvin}$, that is $h_{\mathrm{EB}}=\left.H_{\mathrm{EB}}\middle/H_{\mathrm{EB,max}}\right.$. Equation~\ref{eq:maf} has a stable solution in thermal equilibrium that is equal to $n_{\mathrm{AF}}=\tanh\left(\mathcal{J}_{\mathrm{ex}}A\middle/k_{\mathrm{B}}T\right)$. The approximate time it takes to reach this value is also known from Eq.~\ref{eq:maf} to be $\mathrel{\mytilde}\left(\left.1\middle/\tau_{\downarrow\uparrow}\right.-\left.1\middle/\tau_{\uparrow\downarrow}\right.\right)^{-1}$, with $T=T_{\mathrm{amb}}$. Note that for $t_{\mathrm{AF}}=\SI{10}{\nano\meter}$ it only takes a grain of modest \SI{50}{\square\nano\meter} area for this time to exceed the age of the universe (\SI{4.3e17}{\second}). Therefore, all but the smallest grains will never reach full equilibrium after laser excitation, but rather they stall at some intermediate value between $n_{\mathrm{AF}}=0$ and thermal equilibrium. This is exemplified by the black line in Fig.~\ref{fig:2}d which halts its evolution after $T$ drops below a certain freezing point ((iv) in Fig.~\ref{fig:2}d). More details and equations from this part of the model are found in Appendix~\ref{app:arrhenius}.

\section{Model results and discussion}
\begin{figure}%
	\centering%
	\includegraphics[width=\columnwidth]{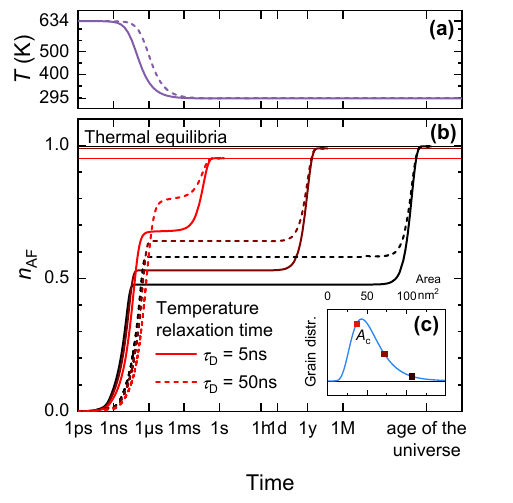}%
	\caption{\label{fig:3}(a) The temperature $T$ after laser excitation according to Eq.~\ref{eq:heatkernel} for $\tau_{\mathrm{D}}=\SI{5}{\nano\second}$ (solid line) and $\tau_{\mathrm{D}}=\SI{50}{\nano\second}$ (dashed line). Time is measured from the moment that $T$ drops below the N\'{e}el temperature $T_{\mathrm{N}}=\SI{634}{\kelvin}$ of \ch{IrMn}. (b) Solutions to Eq.~\ref{eq:maf} of the parameter $n_{\mathrm{AF}}$ for three different grain sizes (see the inset (c) for a distribution of the grain areas and the sampling of the three grain sizes). The smallest grain (red) is taken to be the critical grain area $A_{\mathrm{c}}$ for an antiferromagnet that is $t_{\mathrm{AF}}=\SI{5}{\nano\meter}$ thick. The solid lines are simulated for a temperature relaxation time of $\tau_{\mathrm{D}}=\SI{5}{\nano\second}$ which is typical for the stacks used in the experiment. The dashed lines are simulated for an artificial but achievable value of $\tau_{\mathrm{D}}=\SI{50}{\nano\second}$. The solid horizontal lines at the top of (b) are the steady-state solutions of Eq.~\ref{eq:maf} that represent thermal equilibrium.}%
\end{figure}%
We now use our model to investigate the temporal evolution of individual grains and look at parametric dependencies. For $P_{0}=\SI{58.3e8}{\joule\per\cubic\meter}$ and $t_{\mathrm{AF}}=\SI{5}{\nano\meter}$, the temporal evolution of $T$ and $n_{\mathrm{AF}}$ obtained by solving Eq.~\ref{eq:maf} is shown for three different grain areas in Figs.~\ref{fig:3}a and \ref{fig:3}b. The three grain areas are drawn from the log-normal distribution plotted in Fig.~\ref{fig:2}e with typical values $\mu=\pi\left(\SI{8}{\nano\meter}\right)^{2}$ and $\sigma=0.4$\cite{Khamtawi2023}. The smallest area is chosen to be equal to the critical grain area $A_{\mathrm{c}}$ from Eq.~\ref{eq:ac}, below which grains are superparamagnetic and do not contribute to exchange bias. As is evident from Fig.~\ref{fig:3}, we may distinguish two stages in the setting process: once halting at some intermediate level when $T$ drops below the freezing point and once when reaching thermal equilibrium. For each of the three grain sizes the intermediate level after freezing is slightly different, with smaller grains reaching higher intermediate values of $n_{\mathrm{AF}}$. Moreover, the timescale at which the level is reached is slightly shorter for larger grains. Crucially however, the smaller grains reach thermal equilibrium orders of magnitude earlier than the larger grains. This explains the creep-like evolution of exchange bias observed in our experiments from Fig.~\ref{fig:1}.

\begin{figure}%
	\centering%
	\includegraphics[width=\columnwidth]{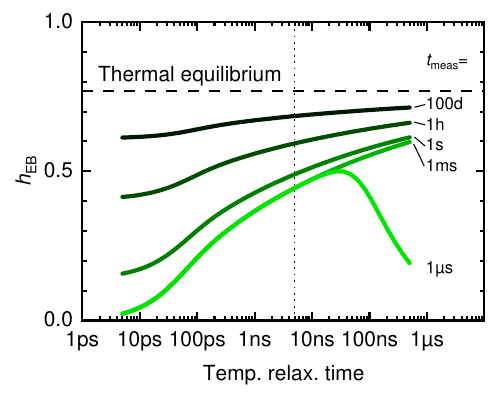}%
	\caption{\label{fig:4} The normalized value of the exchange bias field $h_{\mathrm{EB}}$ reached at various time points $t_{\mathrm{meas}}$ after excitation, plotted as a function of the temperature relaxation timescale $\tau_{\mathrm{D}}$. The values are found by integrating the state of $n_{\mathrm{AF}}$ from the plots in Fig.~\ref{fig:3} over the entire grain size distribution. The values are normalized to the maximum attainable value $H_{\mathrm{EB,max}}$ corresponding to the limiting case of an infinitely thick antiferromagnet at $T=\SI{0}{\kelvin}$. The vertical dotted line corresponds to the expected value $\tau_{\mathrm{D}}=\SI{5}{\nano\second}$ for our stacks. The horizontal dashed line is the thermal equilibrium value of $h_{\mathrm{EB}}$ at room temperature according to Eq.~\ref{eq:hebeq}.}%
\end{figure}%
It is also shown in Fig.~\ref{fig:3} that artificially slowing down the temperature cooling helps improve the freezing value of $n_{\mathrm{AF}}$. The cooling time is determined by the temperature relaxation time $\tau_{\mathrm{D}}$, which enters the model through Eq.~\ref{eq:heatkernel} and is mostly determined by the stack configuration. The solid lines in Fig.~\ref{fig:3} correspond to a typical value for our stacks of $\tau_{\mathrm{D}}=\SI{5}{\nano\second}$, while the dashed lines correspond to a slower relaxation time $\tau_{\mathrm{D}}=\SI{50}{\nano\second}$. It can be seen that for all grains, the freezing value of $n_{\mathrm{AF}}$ is improved by increasing $\tau_{\mathrm{D}}$, even though the time needed to reach thermal equilibrium is not significantly changed.

To make this more quantitative, we simulate the outcome of $h_{\mathrm{EB}}$ for different values of $\tau_{\mathrm{D}}$ and for different measurement times $t_{\mathrm{meas}}$. Figure~\ref{fig:4} depicts the trends of $h_{\mathrm{EB}}$ as a function of $\tau_{\mathrm{D}}$, where $\tau_{\mathrm{D}}=\SI{5}{\nano\second}$ is indicated with a vertical dotted line. From Fig.~\ref{fig:4} we can learn three important things. Firstly, the benefits of a longer $\tau_{\mathrm{D}}$ are always present, but are most pronounced for shorter $t_{\mathrm{meas}}$. Secondly, for very fast measurement \textendash{} i.e., short $t_{\mathrm{meas}}$ \textendash{} there is an optimal $\tau_{\mathrm{D}}$ that maximizes the attainable exchange bias magnitude. Lastly, it can be seen that decreasing $t_{\mathrm{meas}}$ from \SI{1}{\milli\second} to \SI{1}{\micro\second} does not incur any significant losses in the reached value of $h_{\mathrm{EB}}$, which is promising for applications requiring high repetition rates. In practice, manipulation of $\tau_{\mathrm{D}}$ may be achieved by engineering the stack composition or geometry, for example by incorporating thermal insulators that trap the heat in the magnetic layers. Our simulations support this strategy and predict an improvement of \mytilde\num{10} percent points in the magnitude of the reversed exchange bias.

\begin{figure}%
	\centering%
	\includegraphics[width=\columnwidth]{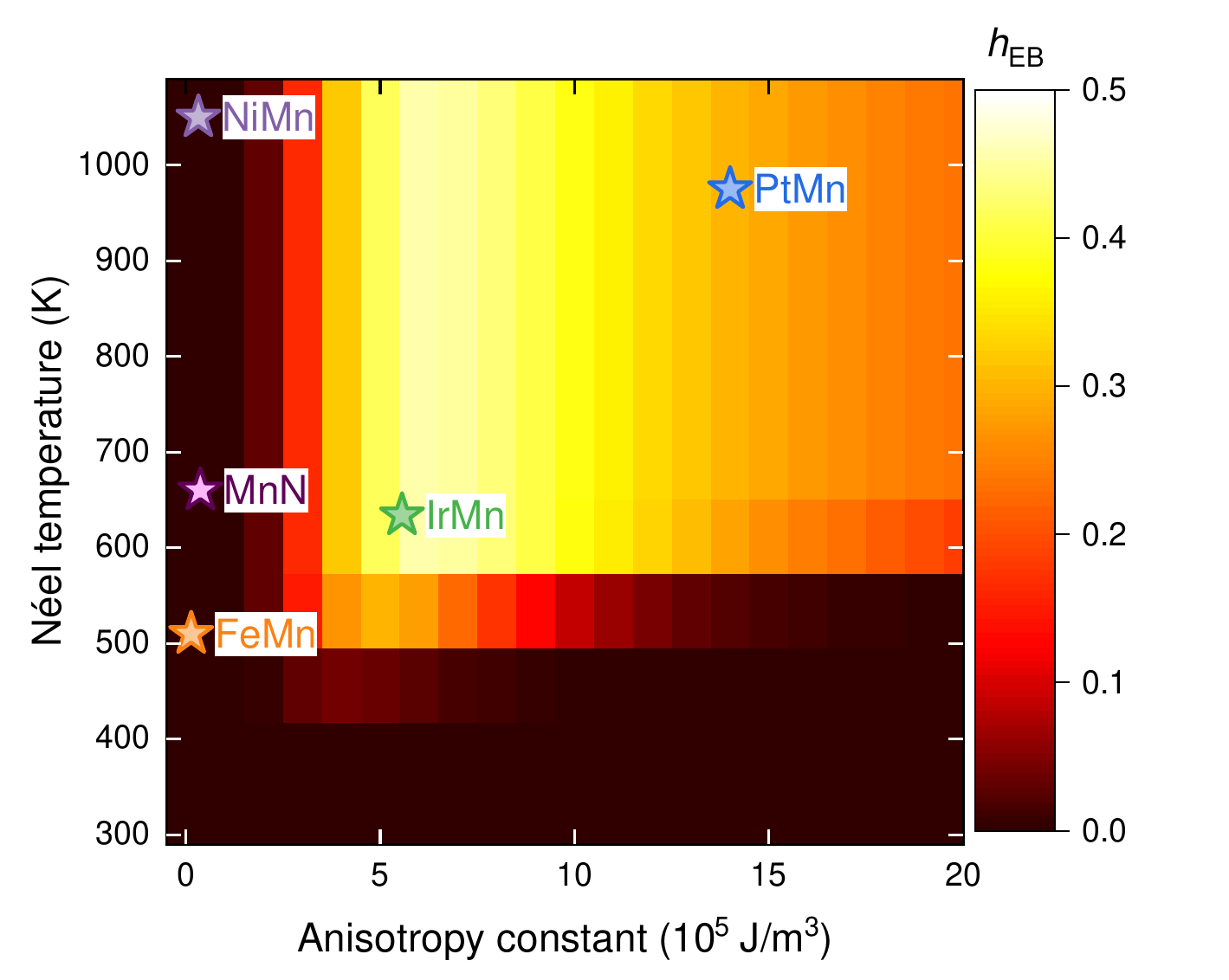}%
	\caption{\label{fig:5}The value of $h_{\mathrm{EB}}$ integrated over all grain areas \SI{1}{\milli\second} after laser excitation for different combinations of N\'{e}el temperature $T_{\mathrm{N}}$ and anisotropy constant $K_{\mathrm{AF},0}$. The stars indicate a selection of examples from real-world materials: \ch{IrMn} \cite{Coey2001,Vallejo-Fernandez2007,Khamtawi2023}, \ch{PtMn} \cite{Coey2001,Umetsu2006}, \ch{NiMn} \cite{Coey2001,Vallejo-Fernandez2007,Umetsu2006}, \ch{FeMn} \cite{Coey2001,Vallejo-Fernandez2007} and \ch{MnN} \cite{Meinert2015}.}%
\end{figure}%

Other parameters besides the temperature relaxation time may also be engineered, for example by choosing a material for the antiferromagnet. Many classes of antiferromagnets exist, with a variety of different N\'{e}el temperatures and anisotropy constants. The \ch{Mn}-based alloys, such as \ch{IrMn} used in our experiments but also \ch{PtMn} or \ch{FeMn}, are conducting and therefore useful in current-perpendicular-to-plane devices like tunnel magnetoresistive sensors. Choosing the parameters $T_{\mathrm{N}}$ and $K_{\mathrm{AF}}$ according to values found for these materials in literature\cite{Coey2001,Vallejo-Fernandez2007,Khamtawi2023,Umetsu2006,Meinert2015}, we can make a reasonable guess for which material is most promising for the type of ultrafast reversal experiment we are conducting. In Fig.~\ref{fig:5}, we show a state diagram for $h_{\mathrm{EB}}$ after \SI{1}{\milli\second} keeping the same parameters as before but varying $T_{\mathrm{N}}$ and $K_{\mathrm{AF}}$, also pointing out where various materials are expected to reside. In fact, our model predicts that \ch{IrMn} occupies a sweet spot in the parameter space, having an anisotropy constant that is large enough to ensure preservation of exchange bias during the rapid temperature settling (unlike \ch{FeMn} or \ch{NiMn}), and small enough that it does not hinder the ability of the grains to transition to the reversed state (unlike \ch{PtMn}).

One way that \ch{IrMn} may be further optimized is by considering its grain size distribution. In \supsec{2} we show how the value of $h_{\mathrm{EB}}$ after \SI{1}{\milli\second} depends on $\mu$ and $\sigma$ from Eq.~\ref{eq:lognormal}. Although these are much harder to control in experiments, it is known that grains are enlarged by annealing at high temperatures and that the buffer layer is also a contributing factor \cite{Nishioka1996}. It is noteworthy to see in the state diagram in \supfig{1} that there appears to be an optimal median grain size $\mu$ and that better performance is expected for narrower distributions (i.e., lower $\sigma$) around this optimal $\mu$. Thus, our model predicts that manipulating the grain size distribution can certainly lead to improved performance.

As a final note, we emphasize that our model can be used to describe various types of systems and experiments. To demonstrate, we performed a fluence dependency study on the stacks used by Guo et al. \cite{Guo2024}, which consist of a \ch{Gd23Co77} ferrimagnetic alloy grown on top of \ch{IrMn}. The experimental results on alloys are different from our multilayers. For example, the alloys lack the annealed plateau observed in Fig.~\ref{fig:1}c. This difference can be understood by considering the fact that the switching fluence threshold is typically larger for multilayers than for alloys\cite{Beens2019}. In other words: when ramping up the fluence $P_{0}$, in our multilayer stacks $T$ exceeds $T_{\mathrm{N}}$ before switching is achieved, while for alloys switching may be achieved before $T$ exceeds $T_{\mathrm{N}}$. Another difference between alloys and multilayers is that the attained value of $h_{\mathrm{EB}}$ is typically lower for alloys than for multilayers, which can be understood from the reduced fraction of \ch{Co}-\ch{Mn} pairs at the ferrimagnetic-antiferromagnetic interface and the negligible \ch{Gd}-\ch{Mn} coupling that does not contribute significantly to the exchange bias setting \cite{Ali2008}. In \supsec{3} we give details on how we implemented the \ch{GdCo} alloys into our model and show that it produces good correspondence between our simulated results and experimental results from Guo et al.\cite{Guo2024}.

This exemplifies that for future research, the versatility of our model can prove useful for predicting the outcomes in a wide variety of situations. For example, the effect of multiple laser pulses can easily be implemented by modifying the temperature evolution in Eq.~\ref{eq:heatkernel}. Moreover, magnetization dynamics under the influence of an external magnetic field may be implemented by a coupled Landau-Lifshitz-Gilbert system.

\section{Conclusions}
In conclusion, we developed a model describing exchange bias reversal on both the ultrashort and long timescales. The model explains observations from field-free ultrafast laser-induced exchange bias reversal experiments, where slow creep-like evolution of the exchange bias occurred at timescales much longer than the ultrashort laser excitation. Heat-driven transitions in antiferromagnetic grains were identified as an underlying mechanism for this effect. We showed that the model is able to predict the reversal fraction and detailed temporal evolution, allowing us to identify parameters that may be tailored for optimizing the performance. In particular, we consider the temperature relaxation timescale a promising candidate and predict an improvement in the magnitude of exchange bias up to \mytilde\num{10} percent points. We also expect multilayer stacks with \ch{IrMn} to outperform any other common antiferromagnetic \ch{Mn}-based alloys, although strategies with different buffer layers or additional annealing steps may be able to push the exchange bias reversal performance even further. Thus, for future research, our modeling framework guides the design and fabrication of optically reprogrammable devices.

\section*{Acknowledgments}
This work was financially supported by the Eindhoven Hendrik Casimir Institute.

\section*{Author contributions}
F.J.F.v.R. wrote the manuscript, performed the experiments and analyzed the data, A.B.M.D. performed and analyzed the long timescale measurements, B.K. and D.C.L. supervised the work and revised the manuscript.

\appendix

\section{Ultrafast magnetization switching}\label{app:m3tm}
The ultrafast part of the model is built up of $N$ macrospin subsystems labeled $i=1,\ldots,N$ that represent, e.g., individual atomic monolayers in a multilayer thin film or a species of atoms in an alloy. Each subsystem has an associated magnetization with $z$-component $m_{i}$ in the range $\left[-1,1\right]$, normalized to its value at $T=\SI{0}{\kelvin}$. In equilibrium, the magnetization $m_{i}$ in layer $i$ is determined by the exchange splitting $\Delta_{i}$ of the spin levels in the layer, i.e., the energy difference between spin levels $s$ and $s\pm1$. We define the matrix $\mathbf{U}$ as%
\begin{eqnarray}%
	U_{ij}=C_{ij}\mathcal{J}_{ij}\frac{\mu_{\mathrm{at},j}}{2\mu_{\mathrm{B}}},\quad\textrm{such that}\nonumber\\%
	\label{eq:delta}\Delta_{i}=2\mu_{\mathrm{B}}\mu_{0}H+\mathbf{U}\cdot\mathbf{m}.%
\end{eqnarray}%
Here, $\mathcal{J}_{ij}$ is the Heisenberg exchange energy between atoms in layers $i$ and $j$, $\mu_{\mathrm{at}}$ is the atomic magnetic moment, $\mu_{B}\approx\SI{9.274}{\joule\per\tesla}$ is the Bohr magneton and $H$ is the externally applied magnetic field. For all our simulations $H=0$, but it may also be set to any non-zero value. Assuming a $(111)$-oriented closely-packed lattice of atoms, we define a coordination matrix $C_{ij}$, which for an alloy $\ch{A}_{x}\ch{B}_{1-x}$ of magnetic elements \ch{A} and \ch{B} takes the form%
\begin{equation}\label{eq:calloy}%
	\mathbf{C}=12\begin{pmatrix}%
		x & & 1-x\\%
		x & & 1-x%
	\end{pmatrix},%
\end{equation}%
while for a multilayer it becomes the $N\times N$ matrix
\begin{equation}\label{eq:cmultilayer}%
	\mathbf{C}=\begin{pmatrix}%
		6 & 3 & 0 & 0 & \dots\\%
		3 & 6 & 3 & 0 & \dots\\%
		0 & 3 & 6 & 3 & \dots\\%
		0 & 0 & 3 & 6 & \dots\\%
		\vdots & \vdots & \vdots & \vdots & \ddots%
	\end{pmatrix}.%
\end{equation}%
Multilayers containing alloys are then easily constructed by mixing Eqs.~\ref{eq:calloy} and \ref{eq:cmultilayer} into a single coordination matrix.

The temperatures $T_{\mathrm{e}}$ and $T_{\mathrm{p}}$ are assumed to be homogeneous across all layers and evolve according to the coupled set of differential equations%
\begin{eqnarray}\label{eq:2tm}%
	\gamma T_{\mathrm{e}}\frac{\mathrm{d}T_{\mathrm{e}}}{\mathrm{d}t}=-g_{\mathrm{ep}}\left(T_{\mathrm{e}}-T_{\mathrm{p}}\right)+\frac{P_{0}}{\Gamma\sqrt{\pi}}\exp\left(-\frac{t^{2}}{\Gamma^{2}}\right), \nonumber\\%
	C_{\mathrm{p}}\frac{\mathrm{d}T_{\mathrm{p}}}{\mathrm{d}t}=g_{\mathrm{ep}}\left(T_{\mathrm{e}}-T_{\mathrm{p}}\right)+C_{\mathrm{p}}\frac{T_{\mathrm{amb}}-T_{\mathrm{p}}}{\tau_{\mathrm{D}}}.%
\end{eqnarray}%
Here, $\gamma=\SI{2.0e3}{\joule\per\cubic\meter\per\kelvin\squared}$ is the temperature coefficient of the electronic heat capacity, $C_{\mathrm{p}}=\SI{4e6}{\joule\per\cubic\meter\per\kelvin}$ is the phonon heat capacity, $g_{\mathrm{ep}}=\SI{4.05e6}{\joule\per\cubic\meter\per\kelvin\per\pico\second}$ is the electron-phonon heat exchange coupling, $\Gamma=\SI{50}{\femto\second}$ is the pulse width, $T_{\mathrm{amb}}=\SI{295}{\kelvin}$ is the ambient temperature and $\tau_{\mathrm{D}}$ is the temperature relaxation time.

An antiferromagnet is modeled in this framework as a layered ferromagnet with $\mathcal{J}_{i,i}>0$ and $\mathcal{J}_{i,i\pm1}<0$. Moreover, the Zeeman splitting term in Eq.~\ref{eq:delta} is ignored for all antiferromagnetic layers.

The magnetization in equilibrium is computed using%
\begin{eqnarray}%
	\label{eq:equilibrium}m_{i}&=&-\frac{1}{S_{i}}\sum_{s=-S_{i}}^{S_{i}}sf_{s,i}~,\\%
	f_{s,i}&=&\left.\exp\left(s\frac{\Delta_{i}}{k_{\mathrm{B}}T_{\mathrm{e}}}\right)\middle/\sum_{s^{\prime}=-S_{i}}^{S_{i}}\exp\left(s^{\prime}\frac{\Delta_{i}}{k_{\mathrm{B}}T_{\mathrm{e}}}\right)\right.,\nonumber%
\end{eqnarray}%
with $f_{s,i}$ the occupation fractions of spin level $s$ in equilibrium following Boltzmann statistics. Equation~\ref{eq:equilibrium} conveniently reduces to $m_{i}=\mathcal{B}_{S_{i}}\left(\left.S_{i}\Delta_{i}\middle/k_{\mathrm{B}}T\right.\right)$ with $\mathcal{B}_{S}$ the Brillouin function for spin quantum number $S$. However, for a full description of the dynamics it is required to keep track of the occupation of all spin levels separately.

The equilibrium values of $f_{s,i}$ are used as the initial conditions for solving a system of ordinary differential equations that describe the evolution of $f_{s,i}$ out of equilibrium. The rate equations consist of one term accounting for Elliot-Yafet spin-flip electron-phonon scattering events (EY) and one term accounting for electron-electron scattering events with exchange of angular momentum (ex),%
\begin{equation}\label{eq:fsi}%
	\frac{\mathrm{d}f_{s,i}}{\mathrm{d}t}=\left.\frac{\mathrm{d}f_{s,i}}{\mathrm{d}t}\right|_{\mathrm{EY}}\left(T_{\mathrm{p}},T_{\mathrm{e}}\right)+\left.\frac{\mathrm{d}f_{s,i}}{\mathrm{d}t}\right|_{\mathrm{ex}}\left(T_{\mathrm{e}}\right).%
\end{equation}%
The full rate equations can be found in Beens et al.\cite{Beens2019}.

The `EY'-term is mostly governed by the material-dependent spin-flip probability $a_{\mathrm{sf}}$. The `ex'-term scales with an integral representing the probability that a scattering event between two electrons exchanging one unit of angular momentum with energy cost $\Delta x=(\Delta_i-\Delta_j)/k_{\mathrm{B}}T_{\mathrm{e}}$ takes place,%
\begin{eqnarray}\label{eq:si}%
	\iiint\mathrm{d}x_{1}\mathrm{d}x_{2}\mathrm{d}x_{3}\,\left[1-f\left(x_{1}+x_{2}+\Delta x-x_{3}\right)\right] \nonumber\\%
	\times\left[1-f\left(x_{3}\right)\right]f\left(x_{2}\right)f\left(x_{1}\right)=\frac{\Delta x}{6}\frac{4\pi^{2}+\Delta x^{2}}{1-\exp\left(-\Delta x\right)}.%
\end{eqnarray}%
The integral in Eq.~\ref{eq:si} was solved analytically for the Fermi-Dirac distribution function $f\left(x\right)=\left.1\middle/\left(e^{x}+1\right)\right.$.

\section{Exchange bias setting}\label{app:arrhenius}
In this part of the model, the evolution of $n_{\mathrm{AF}}$ (Eq.~\ref{eq:maf}) is considered. Since the timescales are much longer than the picosecond rates in the M3TM part, the magnetization $m_{i}$ in each subsystem $i$ of the stack is assumed to be equal to its equilibrium value according to Eq.~\ref{eq:equilibrium}. The temperature is assumed to decay according to a one-dimensional heat kernel with a $t^{-1/2}$ time-dependence. Explicitly, this looks like%
\begin{equation}\label{eq:heatkernel}%
T\left(t\right)=T_{\mathrm{amb}}+\frac{T_{\mathrm{f}}-T_{\mathrm{amb}}}{\sqrt{\left(t-t_{0}\right)/\tau_{\mathrm{D}}+1}}.%
\end{equation}%
Here, $\tau_{\mathrm{D}}$ is the temperature relaxation timescale. $T_{\mathrm{f}}$ and $t_{0}$ are parameters whose explicit formulas have been previously reported \cite{VanRiel2025}.

The critical grain size is determined by the superparamagnetic limit and is defined here as%
\begin{equation}\label{eq:ac}%
A_{\mathrm{c}}=25\frac{k_{\mathrm{B}}T_{\mathrm{amb}}}{K_{\mathrm{AF}}t_{\mathrm{AF}}},%
\end{equation}%
with $T_{\mathrm{amb}}$ the ambient temperature. For calculating $H_{\mathrm{EB}}$, the value of $n_{\mathrm{AF}}$ is integrated over all grain sizes larger than $A_{\mathrm{c}}$, taking the distribution function into account.

The exchange bias at time $t$ is calculated via%
\begin{equation}\label{eq:heb}%
	h_{\mathrm{EB}}\left(t\right)=\int_{A_{\mathrm{c}}}^{\infty}\mathrm{d}\mathbb{L}\,n_{\mathrm{AF}}\left(A,t\right),%
\end{equation}%
where $\mathrm{d}\mathbb{L}$ is the log-normal probability density for the grain size distribution defined by%
\begin{equation}\label{eq:lognormal}%
	\mathrm{d}\mathbb{L}=\frac{1}{A\sigma\sqrt{2\pi}}\exp\left(-\frac{\ln^{2}\left(A/\mu\right)}{2\sigma^{2}}\right)\mathrm{d}A.%
\end{equation}%
The limiting value in thermal equilibrium for a given ambient temperature $T_{\mathrm{amb}}$ is given by%
\begin{equation}\label{eq:hebeq}%
	h_{\mathrm{EB,eq}}=\int_{A_{\mathrm{c}}}^{\infty}\mathrm{d}\mathbb{L}\,\tanh\left(\frac{\mathcal{J}_{\mathrm{ex}}A}{k_{\mathrm{B}}T_{\mathrm{amb}}}\right),%
\end{equation}%
which has been used to calculate the thermal equilibrium value in Fig.~\ref{fig:4}.

\section*{References}
\bibliography{references.bib}

\onecolumngrid
\clearpage

\setcounter{section}{0}
\setcounter{equation}{0}
\setcounter{figure}{0}
\setcounter{table}{0}
\setcounter{page}{1}
\makeatletter
\renewcommand{\appendixname}{Supplementary Section}
\renewcommand{\thesection}{\arabic{section}}
\renewcommand{\theequation}{S\arabic{equation}}
\renewcommand{\thefigure}{S\arabic{figure}}
\renewcommand{\bibnumfmt}[1]{[#1]}
\renewcommand{\citenumfont}[1]{#1}

\section{Dependence of exchange bias coupling on grain area}
To simulate the grain area dependence in more detail, we adapted the exchange bias coupling parameter $\mathcal{J}_{\mathrm{ex},0}$ so that it scales with the grain area according to $\mathcal{J}_{\mathrm{ex},0}=\mathcal{J}_{\mathrm{ex,avg}}\sqrt{A_{0}/A}$ \cite{Malozemoff1987,Uyama1997,Fuke1999}. In order to maintain consistency with experimentally observed values, we use a proportionality constant which makes the average value of $\mathcal{J}_{\mathrm{ex},0}$ for our choice of grain size distribution equal to the measured value of $\mathcal{J}_{\mathrm{ex,avg}}=\SI{0.19}{\milli\joule\per\square\meter}$ reported in literature\cite{Coey2001}. If we enforce this condition the value of $A_{0}$ is fixed according to%
\begin{eqnarray}%
	\mathcal{J}_{\mathrm{ex,avg}}=\frac{\int\mathrm{d}\mathbb{L}\mathcal{J}_{\mathrm{ex},0}A}{\int\mathrm{d}\mathbb{L}A}\nonumber\\%
	\Rightarrow A_{0}=\mu\exp\left(\frac{3}{4}\sigma^{2}\right).%
\end{eqnarray}%
For the values of $\mu$ and $\sigma$ used in the main paper this value becomes $A_{0}=\SI{50.6}{\square\nano\meter}$.

\section{Parameter spaces of exchange bias magnitude}
\begin{figure*}[h]%
	\centering%
	\includegraphics[width=0.5\textwidth]{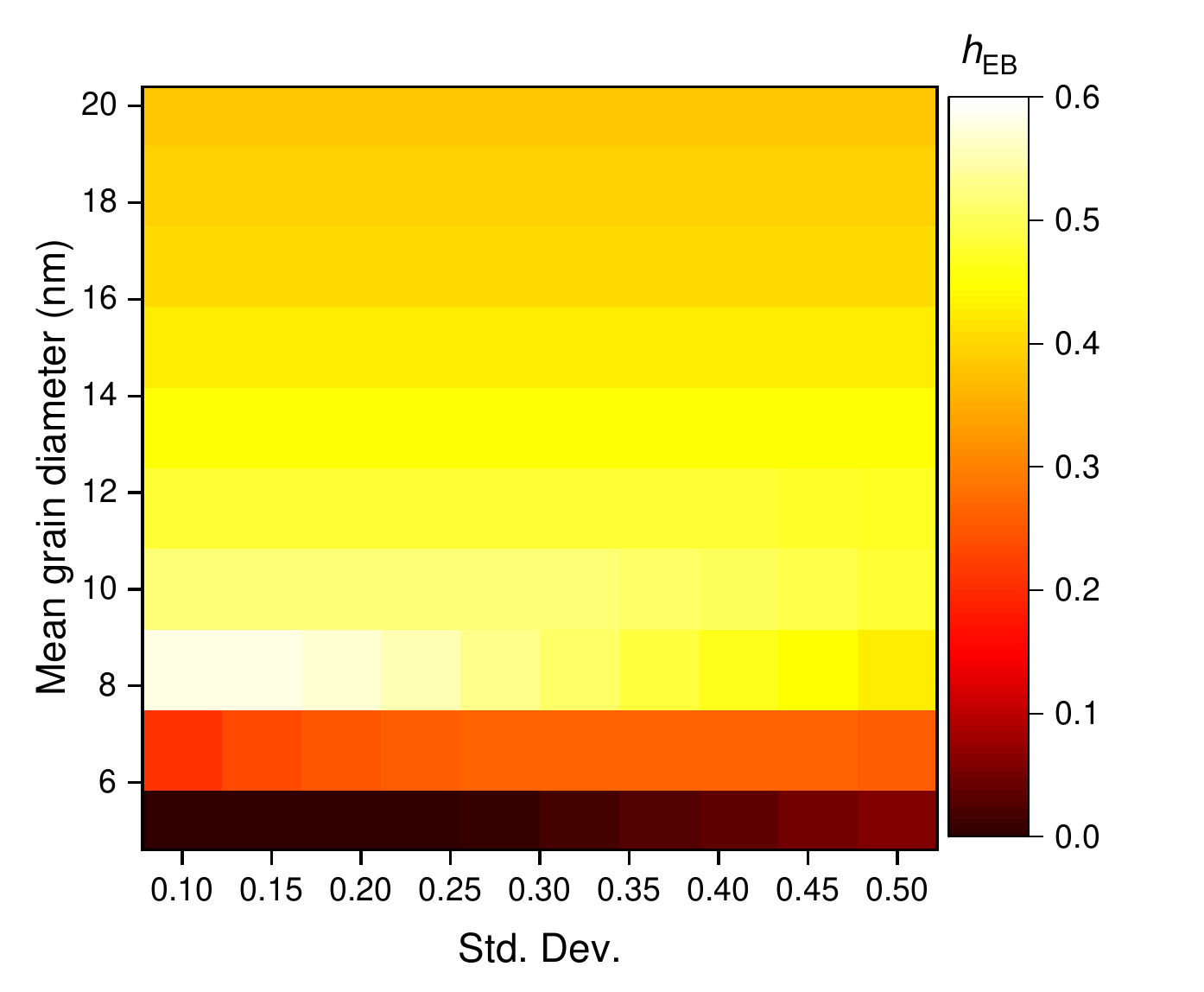}%
	\caption{\label{fig:s1}The value of $h_{\mathrm{EB}}$ integrated over all grain areas \SI{1}{\milli\second} after laser excitation for different combinations of median grain diameter and standard deviation. The modeling results in the paper use a median diameter of \SI{8}{\nano\meter} with a standard deviation of \num{0.4}.}%
\end{figure*}%
We simulated the effects of different grain size distributions in the antiferromagnetic layer on the reversed exchange bias $h_{\mathrm{EB}}$. The results for a \SI{5}{\nano\meter} thick layer are shown in Fig.~\ref{fig:s1}. A clear optimum can be seen in the grain diameter around \SI{8}{\nano\meter}, which is already a typical value found in polycrystalline \ch{IrMn} layers \cite{Khamtawi2023}. However, reducing the standard deviation of the distribution around this optimum does result in an increased $h_{\mathrm{EB}}$ value, getting very close to the thermal equilibrium value at room temperature of \num{0.77}. Thus, attempting to reduce the width of the distribution may lead to increased performance for ultrafast laser-induced exchange bias reversal.

\section{Comparing the model to switching of \ch{IrMn}/\ch{GdCo} from literature}
\begin{figure*}[h]%
	\centering%
	\includegraphics[width=\textwidth]{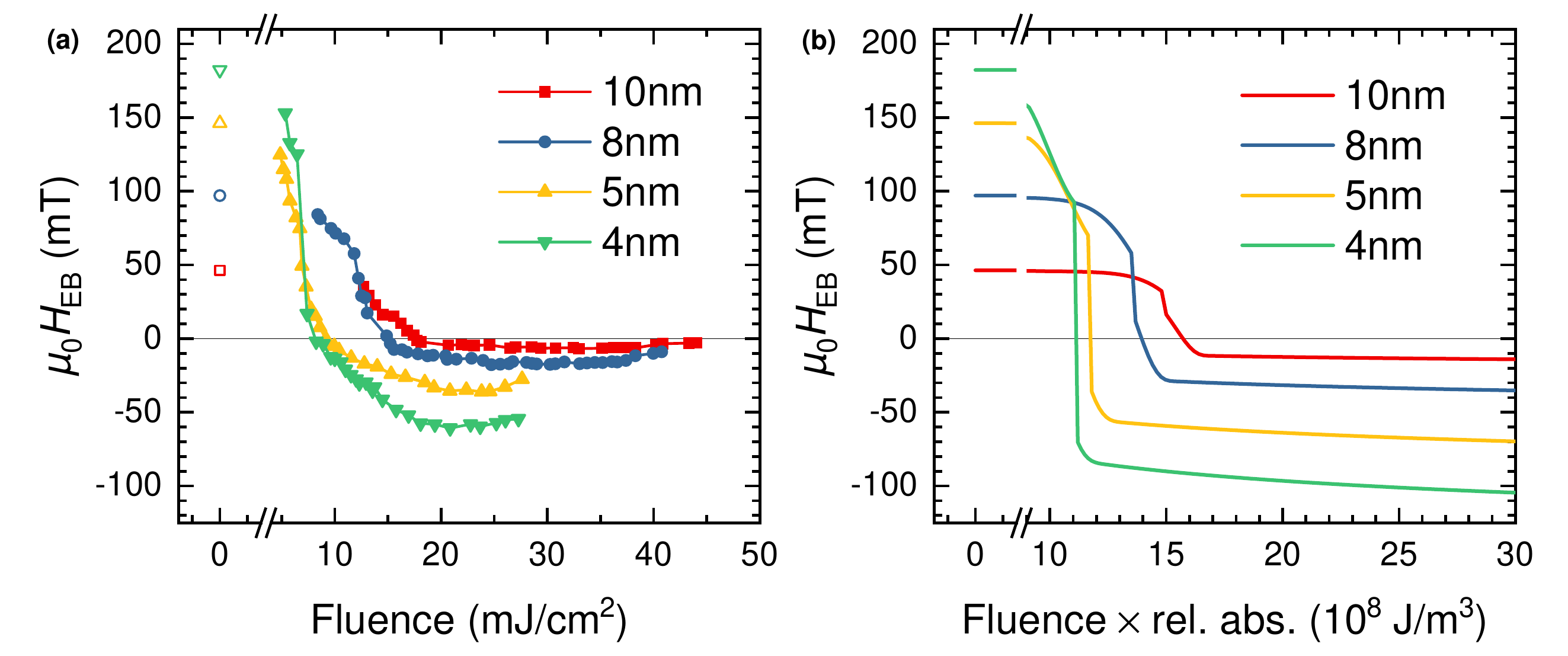}%
	\caption{\label{fig:s2}(a) Experimental results from Guo et al. \cite{Guo2024} of exchange bias magnitude after excitation with a femtosecond laser pulse with varying fluences. The sample consists of a \SI{4}{\nano\meter} thick \ch{Gd_{0.23}Co_{0.77}} ferrimagnet on top of a \ch{IrMn} layer with thicknesses of \SIlist[list-units=single]{4;5;8;10}{\nano\meter}. (b) Simulation results of the model from the main paper, with the value of $h_{\mathrm{EB}}$ scaled according to the values at zero fluence from (a) and with the fluence scaled according to the absorption in the \ch{IrMn} layer relative to the \SI{10}{\nano\meter} thick \ch{IrMn}, as calculated by the transfer matrix method.}%
\end{figure*}%
Here, we attempt to reproduce the fluence dependence of exchange bias reversal in \ch{GdCo} alloys as used in the experiments from Guo et al.\cite{Guo2024} and plotted in Fig.~\ref{fig:s2}a. We model their stack as a \ch{Gd23Co77} alloy adjacent to $4$ \ch{IrMn} layers as was done in the main paper. The coordination matrix of this stack is given by%
\begin{equation}%
	\mathbf{C}=\begin{pmatrix}%
		12x & 12(1-x) & 0 & 0 & 0 & 0 \\%
		12x & 12(1-x) & 0 & 0 & 0 & 0 \\%
		3x  & 3(1-x)  & 6 & 3 & 0 & 0 \\%
		0   & 0       & 3 & 6 & 3 & 0 \\%
		0   & 0       & 0 & 3 & 6 & 3 \\%
		0   & 0       & 0 & 0 & 3 & 6%
	\end{pmatrix},%
\end{equation}%
where $x=0.23$ is the \ch{Gd}-content of the \ch{Gd_{x}Co_{1-x}} alloy. The value for $H_{\mathrm{EB}}$ at zero fluence in Fig.~\ref{fig:s2}a is used as the value for $H_{\mathrm{EB,max}}$ to scale the normalized value of $h_{\mathrm{EB}}$ with. Using these parameters produces the fluence dependence in Fig.~\ref{fig:s2}b, which shows excellent qualitative correspondence with the experimental data in Fig.~\ref{fig:s2}a. Most notably, it can be seen that the annealed plateau is not present because the switching threshold overlaps with the fluence where $T$ exceeds $T_{\mathrm{N}}$, resulting in a sharp jump overlapping the gradual transition of $h_{\mathrm{EB}}$ from positive to negative for increasing fluence. Very similar sharp jumps can be identified in Fig.~\ref{fig:s2}a as well.

In addition, the following small adjustments have been made to the model. We set $\mathcal{J}_{\ch{Gd},\ch{IrMn}}=0$ to reflect the negligible coupling between \ch{Gd} and \ch{IrMn} compared to between \ch{Co} and \ch{IrMn} \cite{Ali2008}. Furthermore, we set $\mathcal{J}_{\mathrm{ex},0}=(1-x)\cdot\SI{0.19}{\milli\joule\per\square\meter}$ to reflect this difference in coupling strength in the exchange bias as well. Finally, we adjust the fluence parameter to the free-space fluence by calculating the absorption in the \ch{IrMn} layer for various antiferromagnetic thicknesses via the transfer matrix method (see Ref.~\citenum{VanRiel2025} and Ref.~\citenum{Polley2023} for details on refractive indices for the individual layers). Normalizing against the absorption in the \SI{10}{\nano\meter} thick layer, this leads to the correction factors \numlist{0.75;0.79;0.91} for \ch{IrMn} thicknesses \SIlist[list-units=single]{4;5;8}{\nano\meter}, respectively. So for example, a \SI{4}{\nano\meter} thick \ch{IrMn} layer only needs \SI{75}{\percent} of the free-space fluence compared to a \SI{10}{\nano\meter} thick layer to achieve the same deposited energy per volume. Moreover, the absorption in the \ch{GdCo} layer is better than in the \ch{IrMn} layer by about a factor \num{1.5}. Therefore we chose a threshold fluence of \SI{15e8}{\joule\per\cubic\meter} for switching the \ch{GdCo} magnetization, which is slightly lower than the threshold fluence predicted by the M3TM (\SI{34e8}{\joule\per\cubic\meter}) but which does best match the experimental results from Guo et al.\cite{Guo2024}. In reality other parameters like $\tau_{\mathrm{D}}$ will also inevitably be influenced by the thickness of \ch{IrMn}, but it would be too cumbersome to attempt implementing all of them. Instead, correcting for the difference in absorption is expected to be enough for capturing most of the effect from varying the \ch{IrMn} thickness.

\end{document}